\documentclass[floats,floatfix,showpacs,amssymb,prd,twocolumn,superscriptaddress,nofootinbib]{revtex4-1}

\newlength{\figw} 
\usepackage{graphicx,epsf, epsfig, amssymb}
\usepackage{bm}
\usepackage{longtable,tabularx}
\usepackage{color}
\usepackage[breaklinks]{hyperref}
\usepackage{amsfonts,amsmath,amssymb,mathrsfs,gensymb}
\usepackage{natbib}
\usepackage{prd_macros}
\usepackage{array}
\usepackage{multirow}
\usepackage{rotating,array}
\usepackage{graphicx}
\newcommand\optional[1]{}
\def\be{\begin{equation}}
\def\ee{\end{equation}}
\def\beq{\begin{eqnarray}}
\def\eeq{\end{eqnarray}}
\usepackage{comment}

\begin{document}

\title{Gravitomagnetic dynamical friction}

\author{Benjamin Cashen}
\email{ben@jwnutritional.com}
\affiliation{Department of Physics, The University of Texas at Dallas, Richardson, TX 75080, USA}

\author{Adam Aker}
\email{axa130531@utdallas.edu}
\affiliation{Department of Physics, The University of Texas at Dallas, Richardson, TX 75080, USA}

\author{Michael Kesden}
\email{kesden@utdallas.edu}
\affiliation{Department of Physics, The University of Texas at Dallas, Richardson, TX 75080, USA}

\pacs{98.10+z, 04.70.Bw, 04.25.Nx}

\date{\today}

\begin{abstract}
A supermassive black hole moving through a field of stars will gravitationally scatter the stars, inducing a backreaction force 
on the black hole known as dynamical friction.  In Newtonian gravity, the axisymmetry of the system about the black hole's
velocity $\mathbf{v}$ implies that the dynamical friction must be anti-parallel to $\mathbf{v}$.  However, in general relativity
the black hole's spin $\mathbf{S}$ need not be parallel to $\mathbf{v}$, breaking the axisymmetry of the system and 
generating a new component of dynamical friction similar to the Lorentz force $\mathbf{F} = q\mathbf{v} \times \mathbf{B}$
experienced by a particle with charge $q$ moving in a magnetic field $\mathbf{B}$.  We call this new force gravitomagnetic
dynamical friction and calculate its magnitude for a spinning black hole moving through a field of stars with Maxwellian
velocity dispersion $\sigma$, assuming that both $v$ and $\sigma$ are much less than the speed of light $c$.  We use
post-Newtonian equations of motion accurate to $\mathcal{O}(v^3/c^3)$ needed to capture the effect of spin-orbit coupling
and also include direct stellar capture by the black hole's event horizon.  Gravitomagnetic dynamical friction will cause a
black hole with uniform speed to spiral about the direction of its spin, similar to a charged particle spiraling about a magnetic
field line, and will exert a torque on a supermassive black hole orbiting a galactic center, causing the angular momentum of
this orbit to slowly precess about the black-hole spin.  As this effect is suppressed by a factor $(\sigma/c)^2$ in
nonrelativistic systems, we expect it to be negligible in most astrophysical contexts but provide this calculation for its
theoretical interest and potential application to relativistic systems.
\end{abstract}

\maketitle 

\section{Introduction} \label{S:Intro}

Chandrasekhar was the first to recognize that a massive perturber moving through a stellar background would scatter stars, 
resulting in a net deceleration in the direction of its motion known as dynamical friction \cite{1943ApJ....97..255C}.
Dynamical friction is responsible for many astrophysical phenomena \cite{2008gady.book.....B} including satellite galaxies
inspiraling towards the centers of their host galaxies
\cite{1980PASJ...32..581M,1982MNRAS.198..707L,1998ApJ...500..575I}, mass segregation of heavy objects in globular
clusters \cite{1969ApJ...158L.139S,1977ApJ...216..883B}, the distribution of galaxies within clusters
\cite{1996ApJ...472..460F,1998ApJ...502..141D}, and the migration of planetesimals in protoplanetary disks
\cite{1999ApJ...513..252O,2004ARA&A..42..549G,2015ApJ...811...54G}.  It also causes supermassive black holes to sink
towards the galactic center following a galaxy merger \cite{1980Natur.287..307B,1996NewA....1...35Q,2001ApJ...563...34M,
2012ApJ...745...83A}, helping them to reach sub-parsec separations at which gravitational radiation is effective in
promoting black-hole mergers.  Dynamical friction also damps out the oscillations of supermassive black holes that have
been displaced from the centers of their host galaxies following a gravitational recoil
\cite{2007PhRvL..98w1101G,2007PhRvL..98w1102C,2007ApJ...659L...5C} produced during a black-hole merger
\cite{2008ApJ...678..780G}.

Dynamical friction can also be interpreted as the gravitational drag force exerted by the effective wake of stars created
behind a massive perturber by its gravitational scattering
\cite{1957MNRAS.117...50D,1972ASSL...31...13K,1983A&A...117....9M}.  If the stellar background is homogeneous and
isotropic in its rest frame, and the perturber is treated as a Newtonian point mass, the system is axisymmetric about the
velocity $\mathbf{v}$ of the perturber and the wake produced will be axisymmetric as well.
However, if the perturber is a black hole, it can have a spin $\mathbf{S}$ \cite{1963PhRvL..11..237K} that need not be
parallel to $\mathbf{v}$, breaking the axisymmetry of the system.  Like a paddle dipped over the side of a canoe, the spin
will disturb the wake of the black hole and cause it to deviate from a straight path through the stellar background.  As the
relative velocities $v$ between supermassive black holes and the stars in their host galaxies are usually much less than
the speed of light $c$, and the distances $r$ between the black holes and these stars is much greater than the
gravitational radius $r_g \equiv GM/c^2$ of a black hole of mass $M$, we will work in the post-Newtonian (PN)
approximation \cite{1915SPAW.......831E} in which the equations of motion are expanded in powers of the small parameters
$(v/c)^2$ and $r_g/r$.  We will show that the lowest-order spin-dependent terms in the scattering of stars by a supermassive
black hole induce a dynamical backreaction force on the black hole in the direction given by $\mathbf{v} \times \mathbf{S}$.
In an analogy with the Lorentz force $q\mathbf{v} \times \mathbf{B}$ experienced by a charge $q$ moving through a
magnetic field $\mathbf{B}$, we dub this force {\it gravitomagnetic dynamical friction}. Although individual stars have
a well defined orbital angular momentum $\mathbf{L}$ with respect to the black hole and exert a gravitomagnetic force
proportional to $\mathbf{v} \times \mathbf{L}$ at lower PN order, symmetry implies that the homogeneous stellar background
has zero total orbital angular momentum and thus exerts no net spin-independent gravitomagnetic dynamical friction.

Recent work on the two-body problem in the PN approximation is too extensive to review here, but the lowest-order
spin-dependent terms have been known for over twenty years \cite{1918PhyZ...19..156L,1951RSPSA.209..248P,1979GReGr..11..149B,1985PhRvD..31.1815T,1993PhRvD..47.4183K,1995PhRvD..52..821K}.
The existing literature on relativistic dynamical friction is far more sparse.  Lee \cite{1969ApJ...155..687L} considered how
the lowest-order spin-independent PN effects modify classical dynamical friction, while Syer \cite{1994MNRAS.270..205S}
allowed both the test mass and scattered particles to be relativistic but only included small-angle scattering.  Motivated by
extreme-mass-ratio inspirals of compact objects into accreting supermassive black holes, Barausse
\cite{2007MNRAS.382..826B} calculated the dynamical friction acting on a test mass moving relativistically through a
perfect fluid.  Pike and Rose \cite{2014PhRvE..89e3107P} extended the work of Spitzer \cite{1962pfig.book.....S} on the
dynamical friction on charged particles moving through plasmas to relativistic systems.  To our knowledge, this
paper provides the first calculation of dynamical friction in which the spin of the test mass is taken into account.  Although
the magnitude of the gravitomagnetic contribution to dynamical friction is strongly suppressed compared to the Newtonian
one for supermassive black holes in realistic host galaxies, we present this calculation for its theoretical interest and
potential applicability to relativistic systems.

In Sec.~\ref{S:scatt}, we review how spinning supermassive black holes scatter stars on hyperbolic orbits
in general relativity.  In Sec.~\ref{S:coeff}, we describe how these results can be used to perform a Monte Carlo calculation of
the gravitomagnetic contribution to the coefficient of dynamical friction. In Sec.~\ref{S:res}, we present the results of this
calculation and explore the how the coefficient depends on the black-hole velocity and spin.  A brief summary and
discussion of the potential astrophysical applications of gravitomagnetic dynamical friction are provided in Sec.~\ref{S:disc}.

\section{Post-Newtonian Orbital Precession} \label{S:scatt}

The equations of motion for two Newtonian point particles with masses $m_1$ and $m_2$ and separation $\mathbf{r}$ can
be converted into an effective one-body equation of motion with acceleration
\begin{equation} \label{E:aNewt}
\mathbf{a}_N = -\frac{M}{r^2}\mathbf{\hat{r}}~,
\end{equation}
where $M \equiv m_1 + m_2$ is the total mass and we use units where $G = c = 1$.  In the PN approximation, this
Newtonian acceleration is supplemented by higher-order corrections \cite{1995PhRvD..52..821K}
\begin{align}
\mathbf{a} = \mathbf{a}_N + \mathbf{a}_{1PN} + \mathbf{a}_{SO} + \ldots
\end{align} 
where
\begin{align} 
\mathbf{a}_{1PN} &= -\frac{M}{r^2} \bigg\{ \mathbf{\hat{r}} \left[ (1+ 3\eta)v^2 - 2(2 + \eta) \frac{M}{r} - \frac{3}{2}\eta\dot{r}^2
\right] \notag \\
& \qquad \qquad - 2(2 - \eta)\dot{r} \mathbf{v} \bigg\} \label{E:a1PN}
\end{align}
and
\begin{align} 
\mathbf{a}_{SO} &= \frac{1}{r^3} \bigg\{ 6\mathbf{\hat{r}} \left[ (\mathbf{\hat{r}} \times \mathbf{v}) \cdot \left( 2\mathbf{S} +
\frac{\delta m}{M} \boldsymbol{\Delta} \right) \right] \notag \\
& \qquad \qquad - \left[ \mathbf{v} \times \left( 7\mathbf{S} + 3\frac{\delta m}{M} \boldsymbol{\Delta} \right) \right] \notag \\
& \qquad \qquad + 3\dot{r} \left[ \mathbf{\hat{r}} \times \left( 3\mathbf{S} + \frac{\delta m}{M} \boldsymbol{\Delta} \right) \right]
\bigg\} \label{E:aSO}
\end{align}
where $\mathbf{v}$ is the relative velocity, $\eta \equiv m_1m_2/M^2$ is the symmetric mass ratio, $\mathbf{S} \equiv
\mathbf{S}_1 + \mathbf{S}_2$ is the total spin, $\delta m \equiv m_1 - m_2$ is the mass difference, and
$\boldsymbol{\Delta} \equiv M(\mathbf{S}_2/m_2 - \mathbf{S}_1/m_1)$.  These expressions simplify in the limit that the
supermassive black hole is much more massive than the stars it scatters ($m_1 \gg m_2$) in which case $\eta \to 0$,
$\mathbf{S} \to \mathbf{S}_1$, $\delta m/M \to 1$, $\boldsymbol{\Delta} \to -\mathbf{S}$, and (with the help of some vector
identites)
\begin{equation} \label{E:aSOlim}
\mathbf{a}_{SO} \to -\frac{2M^2}{r^3}\mathbf{v} \times [-\boldsymbol{\chi} + 3(\mathbf{\hat{r}} \cdot \boldsymbol{\chi})
\mathbf{\hat{r}}]
\end{equation}
where $\boldsymbol{\chi} \equiv \mathbf{S}_1/m_1^2$ is the dimensionless spin of the black hole.

\begin{figure} 
\includegraphics[scale = 0.6]{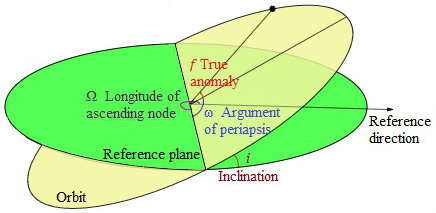}
\caption{A Keplerian orbit in three dimensions, showing the angles that define its orientation in space: the inclination $i$, the argument of periapsis $\omega$, and the longitude of ascending node $\Omega$.  The true anomaly $f$ specifies the
position of a star on its orbit with $f = 0$ at pericenter.} \label{F:OrbEl}
\end{figure}

Dynamical friction is caused by the gravitational scattering of stars by the black hole as it moves through the stellar
background.  These stars are gravitationally unbound to the black hole; they approach from large distances, reach orbital
pericenter, and then return to infinity.  Well before and after each scattering event, the PN accelerations given by
Eqs.~(\ref{E:a1PN}) and (\ref{E:aSO}) are highly subdominant compared to the Newtonian acceleration $\mathbf{a}_N$
and the stellar orbits are well described by Keplerian hyperbolae.  In addition to its semi-major axis $a$ and eccentricity
$e$, a hyperbolic orbit is specified by its inclination $i$, argument of periapsis $\omega$, and longitude of ascending node
$\Omega$ as shown in Fig.~\ref{F:OrbEl}.  The line of nodes is the intersection between a chosen reference plane and
the orbital plane; the portion of this line extending from the origin through the point at which the star crosses the reference
plane from below is called the line of ascending nodes.  The argument of periapsis $\omega$ is the angle in the orbital
plane between the line of ascending nodes and pericenter, while the longitude of ascending node $\Omega$ is the angle in
the reference plane between a chosen reference direction and the line of ascending nodes.  We choose the reference plane
to correspond to the equatorial plane perpendicular to the spin of the black hole implying that the inclination $i$ is the angle
between the orbital angular momentum of the star and the spin of the black hole.  

\begin{figure}[t!]
\includegraphics[scale = 0.3]{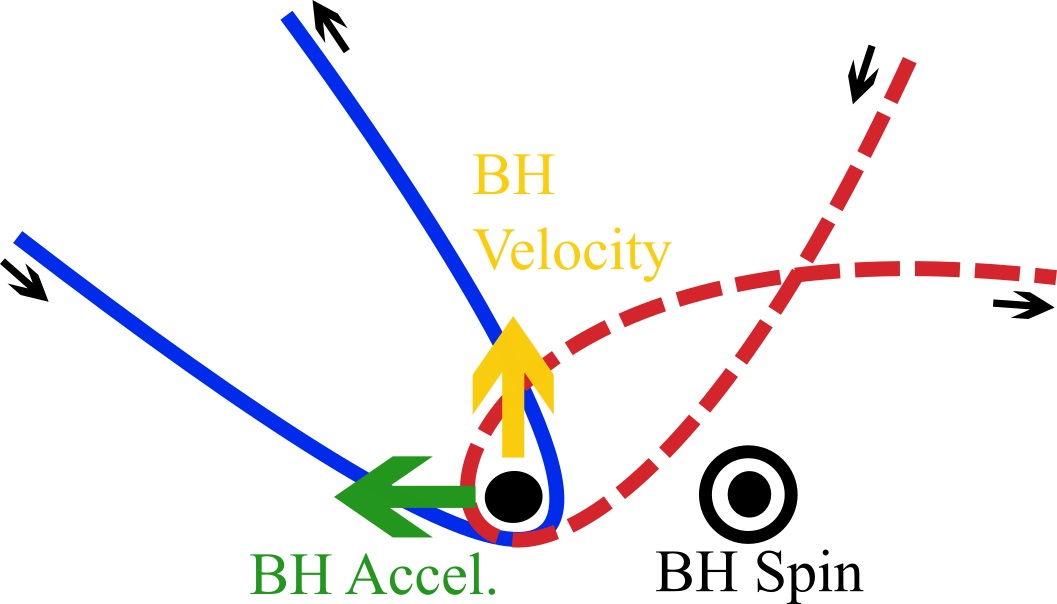}
\caption{Gravitomagnetic dynamical friction caused by spin-dependent precession of the argument of periapsis
$\Delta\omega_{SO}$.  A black hole with velocity directed towards the top of the page (yellow arrow) will encounter more
stars with relative velocities directed towards the bottom of the page as indicated by the downwards black arrows on both
orbits.  The black-hole spin is directed out of the page.  Eq.~(\ref{E:DelomSO}) implies that stars on prograde orbits
($\cos i > 0$, solid blue curve) will experience negative spin-dependent pericenter precession and thus a smaller deflection
than stars on retrograde orbits ($\cos i < 0$, dashed red curve).  The vector sum of the two outwards directed arrows on
these orbits has a component directed to the right side of the page yielding a net backreaction force on the black hole
directed to the left (leftwards green arrow).} \label{F:DelomGDF}
\end{figure} 

Although the five orbital elements $a$, $e$, $i$, $\omega$, and $\Omega$ are conserved by the Newtonian acceleration
$\mathbf{a}_N$, in the Kerr geometry of a spinning black hole, conservation of the energy $E$, $z$-component of the
angular momentum $L_z$, and Carter constant $Q$ \cite{1971PhRvL..26..331C} only require that the first three orbital
elements remain unchanged by the scattering event.  The 1PN acceleration $\mathbf{a}_{1PN}$ given by Eq.~(\ref{E:a1PN})
causes the argument of periapsis for a star on an unbound orbit with specific orbital angular momentum $L$ (orbital
angular momentum divided by the reduced mass) to precess by an amount
\begin{equation} \label{E:Delom1PN}
\Delta\omega_{1PN} = 6\pi \left( \frac{M}{L} \right)^2
\end{equation}
as was famously shown by Einstein to account for the anomalous precession of Mercury's orbit \cite{1915SPAW.......831E}.
Although this pericenter precession should induce a small correction to non-relativistic calculations of dynamical friction, it
does not depend on the black-hole spin and thus cannot generate the gravitomagnetic dynamical friction that is our
concern in this paper.

The 1.5PN spin-orbit correction to the acceleration $\mathbf{a}_{SO}$ causes both the argument of periapsis and the
longitude of ascending node to precess by \cite{2013degn.book.....M}
\begin{subequations} \label{E:SO}
\begin{align}
\Delta\omega_{SO} &= -12\pi\chi\cos i \left( \frac{M}{L}\right)^3~, \label{E:DelomSO} \\
\Delta\Omega_{SO} &= 4\pi\chi\left( \frac{M}{L}\right)^3~. \label{E:DelOmSO}
\end{align}
\end{subequations}
These two changes to the orbital elements during a scattering event each contribute to gravitomagnetic dynamical friction.
Let us first consider the effects of spin-dependent pericenter precession as illustrated in Fig.~\ref{F:DelomGDF} where the
velocity of the black hole is directed towards the top of the page and the black-hole spin is directed out of the page.  The
negative sign in Eq.~(\ref{E:DelomSO}) implies that stars passing on the left side of the black hole (prograde orbits with
$\cos i > 0$) will be scattered by smaller angles than those passing on the right side (retrograde orbits with
$\cos i < 0$).  The black hole will therefore preferentially scatter stars towards the right side of the page and experience a
backreaction force to the left.  This force is anti-aligned with $\mathbf{v} \times \mathbf{S}$ allowing us to
identify it as gravitomagnetic dynamical friction.

\begin{figure}[t!] 
\includegraphics[scale = 0.3]{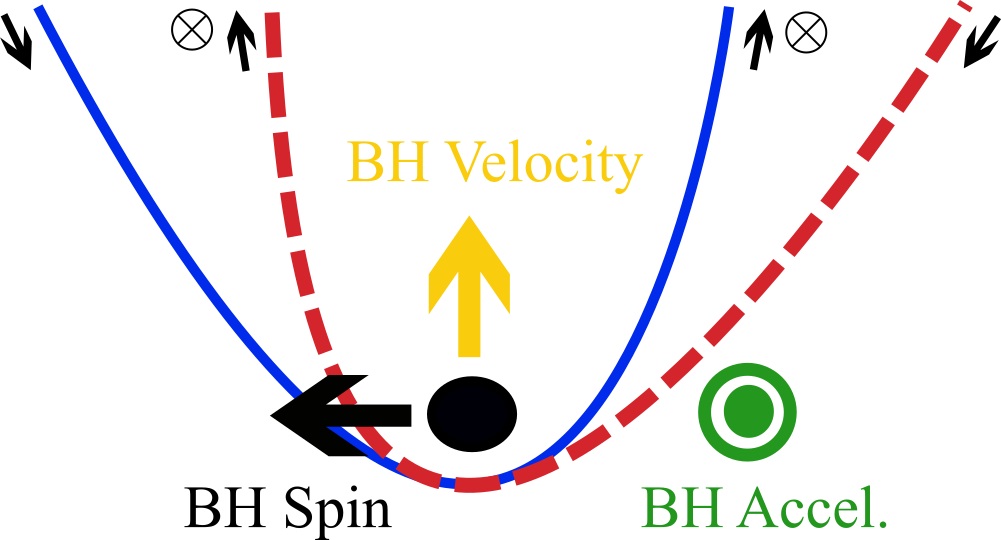}
\caption{Gravitomagnetic dynamical friction caused by precession of the longitude of ascending node
$\Delta\Omega_{SO}$.  The black-hole velocity (yellow arrow) is directed towards the top of the page as in
Fig.~\ref{F:DelomGDF}, but the black-hole spin (black arrow) is now directed towards the left of the page making the solid
blue and dashed red curves correspond to polar orbits with orbital angular momenta directed out of and into the page
respectively.  Precession of the longitude of ascending node causes the orbital angular momenta to precess about the
black-hole spin, deflecting stars on both orbits into the page as indicated by the $\otimes$ symbols on these orbits as they
recede from the black hole.  The backreaction force on the black hole is directed out of the page as shown by the green
$\odot$ symbol.} \label{F:DelOmGDF}
\end{figure}

Let us next consider the effects of precession of the longitude of ascending node as illustrated in Fig.~\ref{F:DelOmGDF}.
The velocity of the black hole is again directed towards the top of the page, but the black-hole spin is now directed to the
left side of the page.  This choice allows the plane of the page to correspond to the initial orbital plane of polar orbits.  The
solid blue and dashed red orbits have angular momenta directed out of and into the page respectively, which according to
Eq.~(\ref{E:DelOmSO}) will precess about the black-hole spin as stars are scattered.  In both cases, the final orbital plane
will dip into the top of the page causing stars to be preferentially scattered below the plane of the page.  This induces a
backreaction force on the black hole out of the page in the direction of $\mathbf{v} \times \mathbf{S}$ as expected for a
gravitomagnetic force.  As this contribution to the gravitomagnetic dynamical friction has the opposite sign of that from the
spin-dependent pericenter precession as shown in Fig.~\ref{F:DelomGDF}, a quantitative calculation is needed to determine
which effect predominates.  We will provide such a calculation in Sec.~\ref{S:coeff}.

We conclude this section by briefly considering relativistic corrections beyond 1.5 PN order.  Black-hole spin induces a
mass-quadrupole moment \cite{1970JMP....11.2580G,1974JMP....15...46H} which in turn generates precession of the
pericenter and longitude of ascending node at 2PN order \cite{2013degn.book.....M}
\begin{subequations}
\begin{align}
\Delta\omega_Q &= -\frac{3\pi\chi^2}{2}(1 - 5\cos^2 i) \left( \frac{M}{L}\right)^4~, \label{E:DelomQ} \\
\Delta\Omega_Q &= 3\pi\chi^2\cos i \left( \frac{M}{L}\right)^4~. \label{E:DelOmQ}
\end{align}
\end{subequations}
These terms have the opposite parity under reflection through the equatorial plane ($\cos i \to -\cos i$) of the 1.5 PN terms
given by Eq.~(\ref{E:SO}) implying that, like the 1PN term of Eq.~(\ref{E:Delom1PN}), they will
not give rise to gravitomagnetic dynamical friction.  We have confirmed this result by showing that including these terms
does not affect our subsequent numerical calculations.

\section{Coefficient of Gravitomagnetic Dynamical Friction} \label{S:coeff}

In this section, we calculate the component of the coefficient $\langle \Delta \mathbf{v} \rangle$ responsible for
gravitomagnetic dynamical friction.  This coefficient specifies the time rate of change of the black hole's velocity, i.e. its
acceleration.  Following Chandrasekhar's original treatment of dynamical friction \cite{1943ApJ....97..255C}, we assume that
the stellar distribution function far from the black hole is spatially homogeneous and has a Maxwellian velocity distribution
\begin{equation} \label{E:Max}
f(\mathbf{x}, \mathbf{v'}) = \frac{n}{(2\pi\sigma^2)^{3/2}} e^{-v'^2/2\sigma^2}~,
\end{equation}
where $n$ is the stellar number density and $\sigma$ is the one-dimensional velocity dispersion.  If the black hole has a
velocity $\mathbf{v'}_{BH}$ in this mean stellar rest frame, we can perform a change of variables $\mathbf{v} \equiv
\mathbf{v'} - \mathbf{v'}_{BH}$ to find a star's velocity in the black-hole rest frame.  In this frame, we can express the
Cartesian components of the stellar velocity $\mathbf{v}$ as
\begin{subequations} \label{E:velvar}
\begin{align}
v_x &= v_r\sin\theta\cos\phi - \frac{L}{r}(\cos\phi_v\sin\phi + \sin\phi_v\cos\theta\cos\phi) \\
v_y &= v_r\sin\theta\sin\phi + \frac{L}{r}(\cos\phi_v\cos\phi - \sin\phi_v\cos\theta\sin\phi) \\
v_z &= v_r\cos\theta + \frac{L}{r}\sin\phi_v\sin\theta
\end{align}
\end{subequations}
where ($\theta, \phi$) is the star's angular position defined such that $\mathbf{v'}_{BH}$ is along the $z$ axis, $v_r$ is
the radial velocity, $L$ is the magnitude of the orbital angular momentum, and $\phi_v$ is an azimuthal angle about the
radial direction defined such that the tangential component of $\mathbf{v}$ is purely azimuthal for $\phi_v = 0$.  In terms
of these variables, the poloidal and azimuthal components of the velocity $\mathbf{v}$ are $v_\theta = -(L/r)\sin\phi_v$ and
$v_\phi = (L/r)\cos\phi_v$.
These new variables, the distribution function (\ref{E:Max}), and the additional definition $\mu \equiv \cos\theta$ allow the
differential flux of stars entering a sphere of radius $r \gg GM/\sigma^2$ to be expressed as
\begin{equation} \label{E:DifFlux}
\frac{d^5F}{d\mu d\phi dv_r dL d\phi_v} = \frac{nLv_r}{(2\pi\sigma^2)^{3/2}}
e^{-|\mathbf{v} + \mathbf{v'}_{BH}|^2/2\sigma^2}~.
\end{equation}
If a star with mass $m_\ast$ scattered by a black hole of mass $M$ experiences a velocity change
$\Delta\mathbf{v}_\ast(\mu, \phi, v_r, L, \phi_v)$, conservation of linear momentum implies that the velocity of the black hole
will be changed by an amount
\begin{equation} \label{E:DeltavBHcon}
\Delta\mathbf{v}_{BH} = -(m_\ast/M)\Delta\mathbf{v}_\ast~.
\end{equation}
Eq.~(\ref{E:DifFlux}) then indicates that the coefficient of dynamical friction $\langle\Delta\mathbf{v}\rangle$ will be
\begin{align}
\langle\Delta\mathbf{v}\rangle &= \int \Delta\mathbf{v}_{BH} \frac{d^5F}{d\mu d\phi dv_r dL d\phi_v}
d\mu\, d\phi\, dv_{r}\, dL\, d\phi_{v} \notag \\
&= -\frac{m_\ast n}{M(2\pi\sigma^2)^{3/2}}e^{-{v'}_{BH}^2/2\sigma^2} \int \Delta\mathbf{v}_\ast Lv_{r} \notag \\
&\quad \times e^{-(v_r^2 + 2v_r v'_{BH} \mu)/2\sigma^2}
d\mu\, d\phi\, dv_{r}\, dL\, d\phi_{v} \label{E:coDF}
\end{align}
where the limits of integration are $-1 < \mu < +1$, $0 < \phi < 2\pi$, $v_r < 0$, $L > 0$, and $0 < \phi_v < 2\pi$.

To evaluate this integral, we must first derive an expression for the change in stellar velocity
$\Delta\mathbf{v}_\ast$.  Qualitatively, there are two possibilities: either the star is directly captured by the black hole's event
horizon or it is scattered by the black hole and returns to infinity.  Let us first consider direct capture by the black hole.  If the
stellar velocity is given by Eq.~(\ref{E:velvar}), the star's specific orbital angular momentum $\mathbf{L} = \mathbf{r} \times
\mathbf{v}$ will be
\begin{subequations} \label{E:Lcart}
\begin{align}
L_x &= L(\sin\phi_v\sin\phi - \cos\phi_v\cos\theta\cos\phi) \\
L_y &= -L(\sin\phi_v\cos\phi + \cos\phi_v\cos\theta\sin\phi) \\
L_z &= L\cos\phi_v\sin\theta~.
\end{align}
\end{subequations}
If we choose without loss of generality for the black-hole spin direction $\mathbf{\hat{S}}$ to lie in the $xz$ plane at an angle
$\theta_S$ with respect to the $z$ axis, the inclination will be
\begin{align}
\cos i &= \mathbf{\hat{L}} \cdot \mathbf{\hat{S}} \notag \\
&= \cos\theta_S\cos\phi_v\sin\theta \notag \\
&\quad + \sin\theta_S(\sin\phi_v\sin\phi - \cos\phi_v\cos\theta\cos\phi)~. \label{E:cosi}
\end{align}
The values $L_{\rm mb\pm}(\chi)$ of the orbital angular momentum below which stars on parabolic orbits in the equatorial
plane of the black hole are captured by the black hole can be found from the geodesic equations; the approximation of
strictly parabolic orbits (specific energy $E = 1$) for both capture and scattering is valid since the star's velocities at infinity
$v \sim \sigma \ll c$ are highly nonrelativistic.  For non-spinning black holes ($\chi = 0$), $L_{\rm mb\pm} = 4M$ since the
inclination is undefined, while for maximally spinning black holes ($\chi = 1$), $L_{\rm mb+} = 2M$ for prograde orbits
($\cos i = +1$) while $L_{\rm mb-} = 2(1 + \sqrt{2})M$ for retrograde orbits ($\cos i = -1$) \cite{1972ApJ...178..347B}.  We can
calculate $L_{\rm mb}$ for non-equatorial orbits by finding the value of $L$ for which there is an unstable circular geodesic
with $E = 1$, $L_z = L\cos i$, and $Q = (L\sin i)^2$; $L_{\rm mb}$ is a monotonically increasing function of the inclination
$i$ with $L_{\rm mb+} \leq L_{\rm mb} \leq L_{\rm mb-}$.  When a star is captured, all of its linear momentum is transferred to
the black hole and $\Delta\mathbf{v}_\ast = -\mathbf{v}$.  Since all stars with $L < L_{\rm mb+}$ are captured, the net linear
momentum they contribute to the black hole must be anti-aligned with $\mathbf{v'}_{BH}$ and thus cannot contribute to the
gravitomagnetic dynamical friction.  The prograde marginally bound orbital angular momentum $L_{\rm mb+}$ can thus
serve as an effective lower limit for the integral in Eq.~(\ref{E:coDF}).

The second possible fate of the star is that it is scattered by the black hole and returns to infinity.  The much larger mass of
the black hole ($M \gg m_\ast$) implies that the star's speed is nearly conserved ($v_i \simeq v_f$).  In Newtonian gravity, a
star with specific energy $\mathcal{E} = E-1$ (not including the rest-mass energy) and specific angular momentum $L$ will
have eccentricity
\begin{equation} \label{E:ecc}
e = \left[ 1 + 2\mathcal{E}\left( \frac{L}{GM} \right)^2 \right]^{1/2}
\end{equation}
and be scattered by an angle
\begin{equation} \label{E:delta}
\delta = 2\cos^{-1}(-e^{-1}) - \pi~,
\end{equation}
where we have temporarily restored the factors of $G$ and $c$ for clarity of presentation.  Stars described by the distribution
function of Eq.~(\ref{E:Max}) have $\mathcal{E} \sim \sigma^2$, and if they are to experience significant spin-orbit coupling
according to Eq.~(\ref{E:SO}) must have $L \gtrsim GM/c$.  This implies eccentricities $e - 1 \simeq (\sigma/c)^2 \ll 1$ and 
deflections $\delta \simeq \pi$.  We can therefore approximate that in the absence of precession, scattering will change a
star's velocity by $\Delta\mathbf{v}_\ast = -2\mathbf{v}$.

Precession will change the direction of the final stellar velocity $\mathbf{v}_f$ so that it is no longer anti-aligned with the
initial velocity $\mathbf{v}$ and thus $\Delta\mathbf{v}_\ast = \mathbf{v}_f - \mathbf{v} \neq -2\mathbf{v}$.  To determine
$\Delta\mathbf{v}_\ast(\mu, \phi, v_r, L, \phi_v)$, we begin by calculating $\mathbf{v}$ and $\mathbf{L}$ from
Eqs.~(\ref{E:velvar}) and (\ref{E:Lcart}) in the limit $r \to \infty$ appropriate for the initial quantities at large separations.  To
determine the components of these vectors in the "spin frame" in which the equatorial plane perpendicular to the black-hole
spin serves as the reference plane shown in Fig.~\ref{F:OrbEl}, we rotate them by an angle $-\theta_S$ about the $y$ axis. 
The stellar inclination $i$ (which remains unchanged by precession) is given by Eq.~(\ref{E:cosi}).  The unit vector
$\mathbf{\hat{n}} = (\mathbf{\hat{S}} \times \mathbf{\hat{L}})/\sin i$ points along the line of ascending node and its
azimuthal angle in spherical coordinates of the spin frame is the initial longitude of ascending node $\Omega$.  Since the
initial stellar velocity $\mathbf{v}$ points towards pericenter for parabolic ($e = 1$) orbits, the argument of periapsis
$\omega$ will be the azimuthal angle of $\mathbf{v}$ in a frame with $\mathbf{\hat{n}}$ along the $x$ axis and
$\mathbf{\hat{L}}$ along the $z$ axis as shown in Fig.~\ref{F:OrbEl}.

Once we have determined the initial orbital elements $\Omega$, $i$, and $\omega$ in the spin frame, the final orbital
elements are
\begin{subequations} \label{E:OrbElFin}
\begin{align}
\Omega_f &= \Omega + \Delta\Omega_{SO}~, \\
i_f &= i~, \\
\omega_f &= \omega + \Delta\omega_{SO}~,
\end{align}
\end{subequations}
where $\Delta\omega_{SO}$ and $\Delta\Omega_{SO}$ are given by Eq.~(\ref{E:SO}).  The final stellar velocity
$\mathbf{v}_f$ in the spin frame is found by successively rotating $-v\mathbf{\hat{x}}$ by the three Euler angles given in
Eq.~(\ref{E:OrbElFin}): we rotate first by $\Omega_f$ about the black-hole spin $\mathbf{S}$, next by $i_f$ about the
line of ascending node $\mathbf{\hat{n}}$, and finally by $\omega_f$ about the orbital angular momentum $\mathbf{L}$.
We then calculate the change in stellar velocity $\Delta\mathbf{v}_\ast = \mathbf{v}_f - \mathbf{v}$ and rotate by an angle
$\theta_S$ about the $y$ axis to transform back from the spin frame to the black-hole rest frame.  We use this
$\Delta\mathbf{v}_\ast$ in Eq.~(\ref{E:coDF}) and Monte Carlo methods to perform the required integration, allowing us to
calculate the coefficient of dynamical friction.  

\section{Results} \label{S:res}

\begin{figure}[t]
\includegraphics[width = \linewidth]{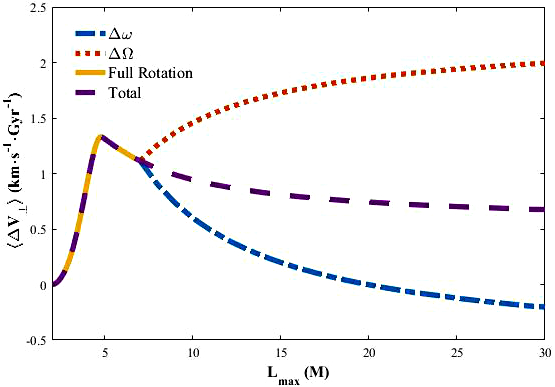}
\caption{The component of the coefficient of dynamical friction $\langle\Delta v_\perp\rangle$ perpendicular to the
black-hole velocity $\mathbf{v}'_{BH}$ as a function of $L_{\rm max}$, the upper limit of the integral over the orbital angular
momentum $L$ in Eq.~(\ref{E:coDF}).   The black hole has a maximal spin ($\chi = 1$) perpendicular to
its velocity ($\theta_s = \pi/2$).  The speed of the black hole is equal to the 1D stellar velocity dispersion
($v'_{BH} = \sigma$) and all other parameters are fixed at the fiducial values listed in the text.  The solid yellow
curve in the range $2M < L < 7M$ shows the total contribution to the gravitomagnetic dynamical friction using the exact Euler
rotations to relate the stellar orbits before and after scattering.  For $L > 7M$ we linearize in the precession angles
$\Delta\omega_{SO}$ and $\Delta\Omega_{SO}$ given by Eq.~(\ref{E:SO}) allowing us to show their separate contributions
with the dot-dashed blue and dotted orange curves respectively.  The dashed purple curve shows the total contribution
which converges for $L_{\rm max} \gtrsim 30M$.} \label{F:DVL}
\end{figure} 

The coefficient of dynamical friction $\langle \Delta\mathbf{v} \rangle$ given by Eq.~(\ref{E:coDF}) is a vector that
consists of two components: the traditional dynamical friction $\langle \Delta v_\parallel \rangle$ anti-aligned with the
black-hole velocity $\mathbf{v}'_{BH}$ and our new coefficient of gravitomagnetic dynamical friction
$\langle \Delta v_\perp \rangle$ perpendicular to $\mathbf{v}'_{BH}$.
In Fig.~\ref{F:DVL}, we show $\langle \Delta v_\perp \rangle$ as a
function of $L_{\rm max}$, the upper limit of the integral over $L$ in Eq.~(\ref{E:coDF}).  We use fiducial parameters
$n = 10^6\,{\rm pc}^{-3}$, $M = 10^6\,M_\odot$, $m_\ast = M_\odot$, and $\sigma = 100\,{\rm km/s}$ typical of a
supermassive black hole moving in a galactic nuclear star cluster.  We measure this acceleration using the unusual units of
km/s/Gyr as galactic speeds and dynamical times are typically measured in km/s and Gyr respectively.  All stars with
$L \leq L_{\rm mb+} = 2M$ are directly captured by the black hole implying by symmetry that they provide zero net
contribution to the gravitomagnetic dynamical friction.  As $L_{\rm max}$ increases from $L_{\rm mb+}$ to $L_{\rm mb-}$,
stars on orbits with increasing inclination manage to escape to infinity and exert a net transverse force on the supermassive
black hole as indicated by the dashed purple curve in Fig.~\ref{F:DVL}.  This curve reaches a maximum at
$L_{\rm max} \simeq 4.875$ quite close to the maximum orbital angular momentum for capture from retrograde orbits
$L_{\rm mb-} = 2(1 + \sqrt{2})$.  The precession angles $\Delta\omega$ and $\Delta\Omega$ diverge for true marginally
bound Kerr geodesics, unlike the 1.5PN results given by Eq.~(\ref{E:SO}), but greater precession would not necessarily lead
to larger gravitomagnetic dynamical friction.  The effect would be eliminated if $\omega_f$ and $\Omega_f$ were fully
randomized.  As it would be computationally prohibitive to solve the geodesic equations for each star in our Monte Carlo
simulation, we rely on the 1.5PN results which should provide a reasonable approximation for values of $L$ modestly above
the marginally bound values.  For $L_{\rm mb+} < L < L_{\rm lin} = 7M$, we use the exact rotations specified by the Euler
angles of Eq.~(\ref{E:SO}) to calculate the velocity change $\Delta\mathbf{v}_\ast$ of each star, but for higher values of $L$
where the precession angles are smaller we linearize in these angles to remove the dominant Newtonian contribution to
dynamical friction and speed up the convergence of our Monte Carlo integration.  This linearization allows us to separate
the contributions from apsidal and nodal precession which partially cancel as anticipated by Figs.~\ref{F:DelomGDF} and
\ref{F:DelOmGDF}.  Our results converge for $L_{\rm max} \gtrsim 30 M$ because the spin-orbit acceleration given by
Eq.~(\ref{E:aSO}) falls off more rapidly than the inverse-square law.

\begin{figure}[t]
\includegraphics[width = \linewidth]{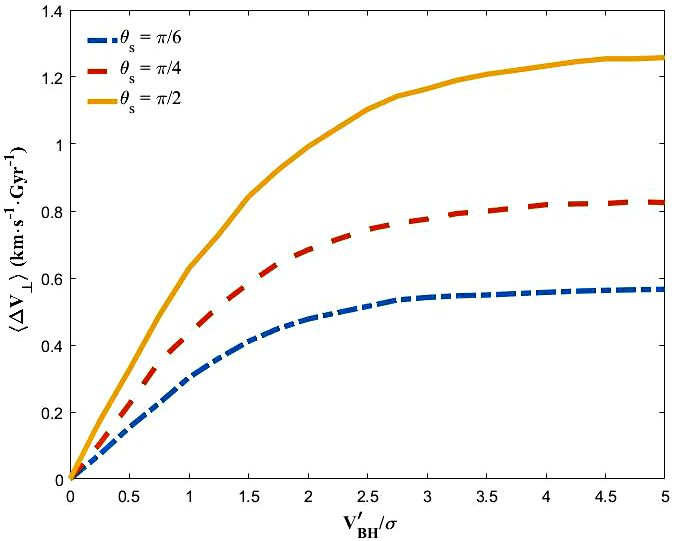}
\caption{The coefficient of gravitomagnetic dynamical friction $\langle \Delta v_\perp \rangle$ as a function of the ratio
$v'_{BH}/\sigma$ between the speed of the black-hole and the stellar velocity dispersion.  Three different values of the
angle $\theta_S$ between the black-hole spin and velocity are shown: $\pi/2$ (solid yellow curve), $\pi/4$ (dashed orange
curve), and $\pi/6$ (dot-dashed blue curve).} \label{F:GDFofV}
\end{figure}

We can obtain a crude estimate for the coefficient of gravitomagnetic dynamical friction $\langle \Delta v_\perp \rangle$ by
assuming that it is comparable to the contribution to the Newtonian coefficient of dynamical friction
$\langle \Delta v_\parallel \rangle$ from stars with orbital angular momenta $L \lesssim L_{\rm max} \sim M$ at which
the spin-orbit precession of Eq.~(\ref{E:SO}) is significant.  In the limit that $v'_{BH},~\sigma \ll c$, we estimate
(using Eq.~(5.20) of Merritt \cite{2013degn.book.....M}) that
\begin{align}
\langle \Delta v_\perp \rangle &\simeq -\frac{2\pi G^2 nMm_\ast}{c^2} \left( \frac{L_{\rm max}}{M} \right) 
K\left( \frac{v'_{BH}}{\sqrt{2} \sigma} \right) \notag \\
&\simeq -1.3~{\rm km/s/Gyr}~\left( \frac{n}{10^6~{\rm pc}^{-3}} \right) \left( \frac{M}{10^6~M_\odot} \right) \notag \\
&\quad \times \left( \frac{m_\ast}{M_\odot} \right) \left( \frac{L_{\rm max}}{M} \right)
K\left( \frac{v'_{BH}}{\sqrt{2} \sigma} \right)~, \label{E:GDFest}
\end{align}
where
\begin{equation} \label{E:Kx}
K(x) \equiv \frac{(2x^2 -1){\rm erf}(x) + x~{\rm erf}'(x)}{2x^2}~,
\end{equation}
and erf and erf$'$ are the error function and its first derivative.  This function has the limiting behavior
$K(x) \simeq 4x/(3\sqrt{\pi})$ for $x \ll 1$ and $K(x) \simeq 1$ for $x \gg 1$.  We show the coefficient of gravitomagnetic
dynamical friction $\langle v_\perp \rangle$ as a function of the ratio $v'_{BH}/\sigma$ in Fig.~\ref{F:GDFofV}.  Our
estimate (\ref{E:GDFest}) provides the correct order of magnitude for $L_{\rm max} \sim M$.  Furthermore,
$\langle v_\perp \rangle$ depends linearly on the black-hole velocity for $v'_{BH} \ll \sigma$ and asymptotes to a constant
value for $v'_{\rm BH} \gg \sigma$ as predicted by Eq.~(\ref{E:Kx}).  

\begin{figure}[t!]
\includegraphics[width = \linewidth]{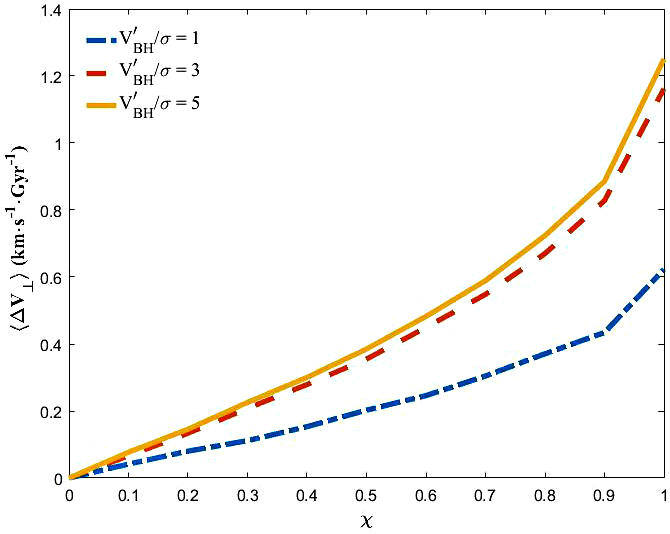}
\caption{The coefficient of dynamical friction $\langle \Delta v_\perp \rangle$ as a function of the dimensionless black-hole
spin $\chi$ for $\theta_S = \pi/2$ and our fiducial parameters $n = 10^6\,{\rm pc}^{-3}$, $m_\ast = M_\odot$, and
$M = 10^6\,M_\odot$.  The three curves show three different choices for the ratio between the black-hole velocity
and stellar velocity dispersion: $v'_{BH}/\sigma = 5$ (solid yellow curve), $v'_{BH}/\sigma = 3$ (dashed red curve), and
$v'_{BH}/\sigma = 1$ (dot-dashed blue curve).} \label{F:GDFofX}
\end{figure}          

We can compare gravitomagnetic dynamical friction to Newtonian dynamical friction for the same stellar distribution function
\begin{align}
\langle \Delta v_\parallel \rangle &= -\frac{4\pi G^2 nMm_\ast \ln\Lambda}{\sigma^2} H\left( \frac{v'_{BH}}{\sqrt{2} \sigma}
\right) \notag \\
&= -2.4 \times 10^7~{\rm km/s/Gyr}~\left( \frac{n}{10^6~{\rm pc}^{-3}} \right) \left( \frac{M}{10^6~M_\odot} \right) \notag \\
&\quad \times \left( \frac{m_\ast}{M_\odot} \right) \left( \frac{\sigma}{100~{\rm km/s}} \right)^{-2}
\ln\Lambda~H\left( \frac{v'_{BH}}{\sqrt{2} \sigma} \right)~, \label{E:NewtDF}
\end{align}
where the Coulomb logarithm $\ln\Lambda \sim 10$ and
\begin{equation} \label{E:Hx}
H(x) \equiv \frac{{\rm erf}(x) - x~{\rm erf}'(x)}{2x^2}
\end{equation}
has the limiting behavior $H(x) \simeq 2x/(3\sqrt{\pi})$ for $x \ll 1$ and $H(x) \simeq 1/(2x^2)$ for $x \gg 1$
\cite{2013degn.book.....M}.  Taking the ratio of Eqs.~(\ref{E:GDFest}) and (\ref{E:NewtDF}), we find
\begin{equation} \label{E:DFrat}
\frac{\langle \Delta v_\perp \rangle}{\langle \Delta v_\parallel \rangle} \simeq \frac{K}{2H\ln\Lambda}
\left( \frac{L_{\rm max}}{M} \right) \left( \frac{\sigma}{c} \right)^2
\end{equation}
which has the limiting behavior
\begin{subequations} \label{E:GDFLim}
\begin{align}
\lim_{v'_{BH} \to 0} \frac{\langle \Delta v_\perp \rangle}{\langle \Delta v_\parallel \rangle} &= \frac{1}{\ln\Lambda} \left( \frac{L_{\rm max}}{M} \right) \left( \frac{\sigma}{c} \right)^2~, \\
\lim_{v'_{BH} \to c} \frac{\langle \Delta v_\perp \rangle}{\langle \Delta v_\parallel \rangle} &= \frac{1}{2\ln\Lambda} \left( \frac{L_{\rm max}}{M} \right) \left( \frac{v'_{BH}}{c} \right)^2~.
\end{align}
\end{subequations}
Although gravitomagnetic dynamical friction is suppressed by a factor $(\sigma/c)^2 \ll 1$ compared to Newtonian
dynamical friction in the nonrelativistic limit, as the black-hole velocity approaches the speed of light these two effects
become comparable.  Although the PN expansion used to calculate these coefficients breaks down in the limit
$v'_{BH} \to c$, there is no reason to expect that higher-order terms will be finely tuned so as to cause a cancellation that
would suppress the gravitomagnetic dynamical friction.

\begin{figure}[t!]
\includegraphics[width = \linewidth]{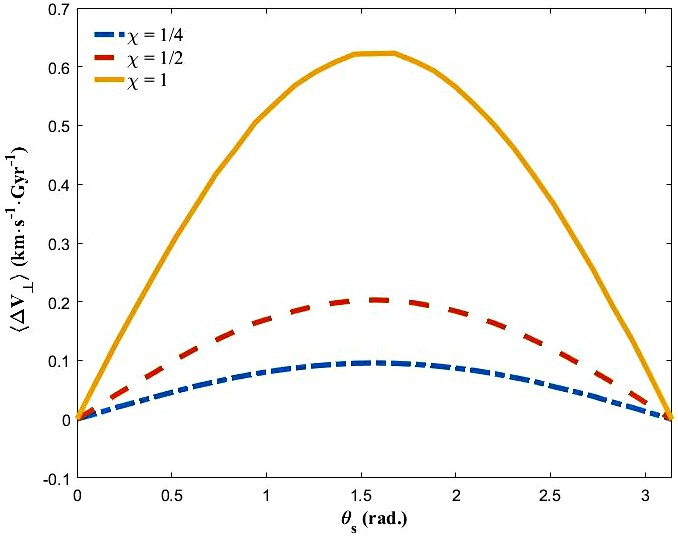} 
\caption{The coefficient of dynamical friction $\langle \Delta v_\perp \rangle$ as a function of the angle $\theta_S$ between
the velocity $\mathbf{v'}_{BH}$ and spin $\mathbf{S}$ of the black hole for the same fiducial parameters listed in the caption
to Fig.~\ref{F:GDFofX}.  The three curves show three different choices for the dimensionless spin magnitude: $\chi = 1$ (solid
yellow curve), $\chi = 1/2$ (dashed red curve), and $\chi = 1/4$ (dot-dashed blue curve).} \label{F:GDFofTheta}
\end{figure}

We show how the coefficient of gravitomagnetic dynamical friction $\langle \Delta v_\perp \rangle$ depends on the
black-hole spin in Figs.~\ref{F:GDFofX} and \ref{F:GDFofTheta}.  Fig.~\ref{F:GDFofX} shows the dependence on the
dimensionless spin magnitude $\chi$ which appears approximately linear for $\chi \lesssim 0.7$ but then grows more
steeply as one approaches the maximal spin limit $\chi = 1$.  Although the precession angles given by Eq.~(\ref{E:SO}) are
linear in the dimensionless spin $\chi$, this linearity is only preserved for contributions to $\langle \Delta v_\perp \rangle$
with $L > L_{\rm lin}$ for which we linearize the Euler rotations in these angles.  The nonlinearity in both these rotations and
the spin dependence of the limits $L_{\rm mb}$ on marginally bound orbits account for the steepening slopes of the
curves $\langle \Delta v_\perp \rangle(\chi)$ seen in Fig.~\ref{F:GDFofX} for $\chi \gtrsim 0.7$.

Fig.~\ref{F:GDFofTheta} shows how the coefficient of gravitomagnetic dynamical friction $\langle \Delta v_\perp \rangle$
depends on the angle $\theta_S$ between the velocity $\mathbf{v'}_{BH}$ and spin $\mathbf{S}$ of the black hole.  The
axisymmetry of the system when $\mathbf{v'}_{BH}$ and $\mathbf{S}$ are parallel ($\theta_S = 0$ or $\pi$) implies that
the coefficient must vanish for these values.  The symmetry of geodesics reflected through the equatorial plane
perpendicular to the black-hole spin implies that the coefficient must be symmetric under reflection through this plane
($\theta_S \to \pi/2 - \theta_S$).  The combination of these two properties suggests that the direction of the gravitomagnetic
dynamical friction force is parallel to $\mathbf{\hat{v}} \times \mathbf{\hat{S}}$ like the Lorentz force law
$\mathbf{F} = q\mathbf{v} \times \mathbf{B}$, even if it is not strictly linear in the magnitudes of $\mathbf{v}$ or $\mathbf{S}$
as indicated by Figs.~\ref{F:GDFofV} and \ref{F:GDFofX}.

Gravitomagnetic friction described by the force law $\mathbf{F} = C\mathbf{v} \times \mathbf{S}$, where
\begin{equation} \label{E:FC}
C = \frac{M\langle \Delta v_\perp \rangle}{vS\sin\theta_S}~,
\end{equation}
would cause a black hole to move on a helical orbit about an axis parallel to its spin $\mathbf{S}$, just as the Lorentz force
law causes a charged particle to move on a helix about the magnetic field.  The radius of this helix would be
\begin{equation} \label{E:HR}
R = \frac{Mv\sin\theta_S}{CS} = \frac{(v\sin\theta_S)^2}{\langle \Delta v_\perp \rangle}~.
\end{equation}
For $v \simeq 100$~km/s and $\langle \Delta v_\perp \rangle$ given by Eq.~(\ref{E:GDFest}), $R \simeq 10$\,Mpc, far larger
than any stellar system with a density as high as $n = 10^6\,{\rm pc}^{-3}$.  A black hole moving on a circular orbit with
orbital angular momentum $\mathbf{L}$ will experience an orbit-averaged torque
\begin{equation}
\frac{d\mathbf{L}}{dt} = -\frac{C}{2M}\mathbf{S} \times \mathbf{L} = \boldsymbol{\Omega}_p \times \mathbf{L}
\end{equation}
implying that $\mathbf{L}$ will precess about $\mathbf{S}$ with a period $\tau_p = 2\pi/\Omega_p \simeq 100$\,Gyr for our
fiducial parameters, far longer than the dynamical-friction time \cite{2013degn.book.....M}
\begin{align}
T_{df} &= \left| \frac{v}{\langle \Delta v_\parallel \rangle} \right| = \frac{3}{8} \sqrt{\frac{2}{\pi}}
\frac{\sigma^3}{G^2Mnm_\ast\ln\Lambda} \notag \\
&\simeq 1.6 \times 10^3\,{\rm yr} \label{E:DFtime}
\end{align}
for our fiducial parameters and $v \ll \sigma$.

\section{Discussion} \label{S:disc}

This paper introduces the concept of gravitomagnetic dynamical friction, a component of dynamical friction perpendicular to
both the velocity and spin of a supermassive black hole traveling through a stellar background.  This force results from the
spin-dependent gravitational scattering of stars on hyperbolic orbits in general relativity.  We calculate the coefficient of
gravitomagnetic dynamical friction $\langle \Delta v_\perp \rangle$ numerically using apsidal and nodal precession at
1.5PN order given by Eq.~(\ref{E:SO}) and the exact spin-dependent angular-momentum threshold for direct capture.  We
also provide a reasonable analytical estimate of this effect in Eq.~(\ref{E:GDFest}) based on the assumption supported by
our Monte-Carlo simulations that gravitomagnetic dynamical friction is produced by scattering stars with $L \sim GM/c$
where relativistic effects are significant.  This estimate suggests that our new coefficient $\langle \Delta v_\perp \rangle$ is
suppressed by a factor $(\sigma/c)^2$ compared to the Newtonian coefficient of dynamical friction
$\langle \Delta v_\parallel \rangle$ anti-aligned with the black-hole velocity $\mathbf{v'}_{BH}$.  This suppression implies
that gravitomagnetic dynamical friction will be negligible in galactic systems where $v'_{BH},\,\sigma \sim~100$\,km/s and
$(\sigma/c)^2 \sim 10^{-7}$.  However, the limiting behavior of our estimate shown by Eq.~(\ref{E:GDFLim}) suggests that as
the black-hole velocity becomes relativistic ($v'_{BH} \to c$), the two coefficients become comparable and therefore our new
effect could be relevant to a black hole interacting with a photon background, circumbinary disk, or relativistic plasma
\cite{1994MNRAS.270..205S,2007MNRAS.382..826B,2014PhRvE..89e3107P}.  Even in the absence of an immediate
application to such a system, it is interesting to consider this qualitatively new dynamical effect introduced by black-hole spin
interacting with a collisionless many-body system in general relativity.

\begin{acknowledgments}
We thank Juan Servin for sharing a program to calculate the angular momentum $L_{\rm mb\pm}$ of marginally bound
orbits as a function of black-hole spin.  M. K. is supported by Alfred P. Sloan Foundation Grant No.~FG-2015-65299 and
NSF Grant No.~PHY-1607031.
\end{acknowledgments}

\bibliography{Feb19_17}
\end{document}